\documentclass[conference,10pt,fleqn]{IEEEtran} 
\usepackage[top=0.5in, bottom=0.75in, left=0.75in, right=0.75in]{geometry}
\usepackage[]{times} 
\usepackage{epsfig} 
\usepackage{subfigure}
\usepackage{amsmath} 
\usepackage{amssymb} %\usepackage{amsthm} 
\usepackage{graphicx} 
\usepackage{subfigure}
\usepackage{multirow}
\usepackage{algorithm2e} 
\renewcommand{\baselinestretch}{1.0} % For 1.5
\newcommand{\bega}{\begin{eqnarray}}
\newcommand{\ega}{\end{eqnarray}}
\newcommand{\bb}{\begin{equation}}
\newcommand{\ee}{\end{equation}}
\newtheorem{defn} {Definition}
\newtheorem{te}{Theorem}
\newtheorem{lema}{Lemma}

\begin{document}
\title{LDPC Codes Which Can Correct Three Errors Under Iterative Decoding}

\author{\authorblockN{Shashi Kiran Chilappagari}
\authorblockA{Dept. of Electrical and Computer Eng.\\
University of Arizona\\
Tucson, AZ 85721, USA \\
Email: shashic@ece.arizona.edu}
\and
\authorblockN{Anantha Raman Krishnan}
\authorblockA{Dept. of Electrical and Computer Eng.\\
University of Arizona\\
Tucson, AZ 85721, USA \\
Email: ananthak@ece.arizona.edu}
\and
\authorblockN{Bane Vasi\'{c}}
\authorblockA{Dept. of ECE and Dept. of Mathematics\\
University of Arizona\\
Tucson, AZ 85721, USA\\
Email: vasic@ece.arizona.edu}}

%\markboth{IEEE Transactions On Information Theory }{Chilappagari \MakeLowercase{\textit{et al.}}: Expander Graph Arguments for Unreliable Memories}

\maketitle
\thispagestyle{empty}
\begin{abstract}
In this paper, we provide necessary and sufficient conditions for a column-weight-three LDPC code to correct three errors when decoded using Gallager A algorithm. We then provide a construction technique which results in a code satisfying the above conditions. We also provide numerical assessment of code performance via simulation results.
\end{abstract}
\section{Introduction}
Iterative message passing algorithms for decoding low-density parity-check (LDPC) codes have been the focus of research over the past decade and most of their properties are well understood \cite{richardsonurbanke},\cite{richardsonurbankeshokrollahi}.  These algorithms operate by passing messages along the edges of a graphical representation of the code known as the Tanner graph and are optimal when the underlying graph is a tree. Message passing decoders perform remarkably well which can be attributed to their ability to correct errors beyond the traditional bounded distance decoding capability. However, in contrast to bounded distance decoders (BDDs), iterative decoders cannot guarantee correction of a fixed number of errors at relatively short code lengths. This is due to the fact that the associated Tanner graphs for short length codes have cycles and the decoding becomes suboptimal and there exist a few low-weight patterns (termed as near codewords \cite{weaknessmackay} or trapping sets \cite{rich}) uncorrectable by the decoder. It is now well established that the trapping sets lead to the phenomenon of error floor. Roughly, error floor is an abrupt change in the frame error  rate (FER) performance of an iterative decoder in the high  signal-to-noise ratio (SNR) region. 

The error floor problem is well understood for iterative decoding over binary erasure channel (BEC) \cite{di}. The decoder fails when the received vector contains erasures in locations corresponding to a stopping set. For the AWGN channel, Richardson in \cite{rich}  presented a  numerical method to estimate error floors of LDPC codes. He established a relation between trapping sets and the FER performance of the code in the error floor region (the necessary definitions will be given in the next section). The approach from \cite{rich} was further refined by Stepanov \textit{et al} in \cite{misha2}. Vontobel and Koetter \cite{koetter} established a theoretical framework for finite length analysis of message passing iterative decoding based on graph covers. This approach was used by Smarandache \textit{et al} in \cite{roxy} to analyze performance of LDPC codes from projective and  for LDPC convolutional codes \cite{roxy2}. For the binary symmetric channel (BSC), error floor estimation based on trapping sets was proposed in \cite{chilappagarione} and we adopt the notation from \cite{chilappagarione}.

In this paper, we make the following two fundamental contributions: (a) give necessary and sufficient conditions for a column-weight-three LDPC code to correct three errors, and (b) propose a construction method which results in a code satisfying the above conditions. 

We consider hard decision decoding for transmission over BSC. The BSC is a simple yet useful channel model used extensively in areas where decoding speed is a major factor. Note that the problem of recovering from a fixed number of erasures is solved for the BEC. If the Tanner graph of a code does not contain any stopping sets up to size $t$ (the size of minimum stopping set is $t+1$), then the decoder is guaranteed to recover from any $t$ erasures. An analogous result for the BSC is still unknown.  The problem of guaranteed error correction capability is known to be difficult and in this paper, we present a first step toward such result. Previously, expansion arguments were used to show that message passing can correct a fixed fraction of errors \cite{burshtein}. However, the code length needed to guarantee such correction capability is generally very large and to correct three errors, the length would be in the order of a few hundred thousand. Also, these arguments cannot be used for column-weight-three codes. Column-weight-three codes are of special importance as their decoders have very low complexity and are used in a wide range of applications.

We also show that the slope of the frame error rate (FER) is dependent on the critical number  of the most relevant trapping sets and hence the slope can be improved by avoiding such trapping sets. We provide a technique to construct codes which outperform empirically best known codes of the same length. Our method can be seen as a modification of the progressive edge growth (PEG) technique proposed in \cite{peg}.

The rest of the paper is organized as follows. In Section \ref{section2} we establish the notation, describe the Gallager A algorithm and define trapping sets. In Section \ref{section3} we present the main theorem which gives the necessary and sufficient conditions to correct three errors. In Section \ref{section4} we describe a technique to construct codes satisfying the conditions of the theorem and provide numerical results. We conclude with a few remarks in Section \ref{section5}
\section{Decoding Algorithms and Trapping Sets}\label{section2}

In this section, we establish the notation and describe a hard decision decoding algorithm known as Gallager A algorithm. We then characterize the failures of the Gallager A decoder with the help of fixed points. We also introduce the notions of trapping sets and critical number.
\subsection{Graphical Representations of LDPC Codes}
The Tanner graph of an LDPC code, $\cal{G}$, is a bipartite graph with two sets of nodes: variable (bit) nodes and check (constraint) nodes. Every edge $e$ in the bipartite graph is associated with a variable node $v$ and check node $c$. The check nodes / variable nodes connected to a variable node / check node are referred to as its neighbors.  The degree of a node is the number of its neighbors. In a $(\gamma,\rho)$ regular LDPC code, each variable node has degree of $\gamma$ and each check node has degree $\rho$. The girth $g$ is the length of the shortest cycle in $\cal{G}$. In this paper, $\bullet$ represents a variable node, $\square$ represents an even degree check node and $\blacksquare$ represents an odd degree check node. 

\subsection{Hard Decision Decoding Algorithms}
Gallager in \cite{gallager} proposed two simple binary message passing algorithms for decoding over the BSC; Gallager A and Gallager B. See \cite{shokrollahi} for a detailed description of Gallager B algorithm. For column-weight-three codes, which are the main focus of this paper, these two algorithms are the same. Every round of message passing (iteration) starts with sending messages from variable nodes (first half of the iteration) and ends by sending messages from check nodes to variable nodes (second half of the iteration). Initially, the variable nodes send their received values to the neighboring checks. In the $k^{th}$ iteration $(k=2,3,\ldots)$, a variable node, $v$ sends the following message, $\stackrel{\textstyle \longrightarrow}{\rm {m_{i}}}(e)$,  along edge $e$ to its neighboring check node $c$ ; if all incoming messages to $v$ other than the message from $c$ are equal to a certain value, it sends that value; else, it sends the received value. A check node $c$ sends to a variable node $v$, the modulo two sum of all incoming messages except the message from $v$. At the end of each iteration, an estimate of each variable node is made based on the incoming messages and possibly the received value. The decoder is run  until a valid codeword is found or for a maximum number of iterations is reached, whichever is earlier. See \cite{jrnlpaperthreeerrors} for a detailed description of the messages passed in Gallager A algorithm.

\textit{A Note on the Decision Rule:} Different rules to estimate a variable node after each iteration are possible and it is likely that changing the rule after certain iterations may be beneficial. However, the analysis of various scenarios is beyond the scope of this paper. For column-weight-three codes only two rules are possible.
\begin{itemize}
\item Decision rule A: if all incoming messages to a variable node from neighboring checks are equal, set the variable node to that value; else set it to received value
\item Decision rule B: set the value of a variable node to the majority of the incoming messages; majority always exists since the column-weight is three 
\end{itemize}
We adopt Decision rule A throughout this paper.
\subsection{Trapping Sets of Gallager A Algorithm}
We now characterize  failures of the Gallager A decoder using fixed points and trapping sets. Much of the following discussion appears in \cite{colwtthreepaper},\cite{jrnlpaperthreeerrors},\cite{chilappagarione},\cite{ucsdpaper} and we include it for sake of completeness. Consider an LDPC code of length $n$ and let $\underline{x}$ be the binary vector which is the input to the Gallager A decoder. Let $\mathcal{S}(\underline{x})$ be the support of $\underline{x}$. The support of $\underline{x}$ is defined as the set of all positions $i$ where $x_i \neq 0$.
\begin{defn}\cite{colwtthreepaper}
A decoder failure is said to have occurred if the output of the decoder is not equal to the transmitted codeword.
\end{defn} 
\begin{defn}\cite{colwtthreepaper}
$\underline{x}$ is called a \textit{fixed point} if for every edge $e$ and its associated variable node $v$
\begin{eqnarray}
\stackrel{\textstyle \longrightarrow}{\rm {m_{k}}}(e) &=& \underline{x}(v), \forall k \nonumber  
\end{eqnarray}
\end{defn}

That is, the message passed from variable nodes to check nodes along the edges are the same in every iteration. Since the outgoing messages from variable nodes are same in every iteration, it follows that the incoming messages from check nodes to variable nodes are also same in every iteration and so is the estimate of a variable after each iteration. In fact, the estimate after each iteration coincides with the received value. It is clear from above definition that if the input to the decoder is a fixed point, then the output of the decoder is the same fixed point. Without loss of generality, we assume that the all zero codeword is sent over BSC and the input to the decoder is the error vector. So, a fixed point with small weight means that few errors lead to decoder failure. A detailed discussion about different kinds of decoder failures is given in \cite{ucsdpaper}
\begin{defn}\cite{chilappagarione}
The support of a fixed point is known as a trapping set. A $(V,C)$ trapping set $\cal{T}$ is a set of $V$ variable nodes whose induced subgraph has $C$ odd degree checks. 
\end{defn}

Our definition of a trapping set gives necessary and sufficient conditions for a set of variable nodes to form a trapping set. We state the following theorem which is a consequence of Fact 3 from \cite{rich}. 
\begin{te}\label{thm1}\cite{colwtthreepaper}
Let $\cal{T}$ be a set consisting of $v$ variable nodes with induced subgraph $\cal{I}$. Let the checks in $\cal{I}$ be partitioned into two disjoint subsets; $\cal{O}$ consisting of checks with odd degree and $\cal{E}$ consisting of checks with even degree. Let $|\mathcal{O}|=c$ and $|\mathcal{E}|=s$.  $\cal{T}$ is a trapping set if : (a) Every variable node in $\cal{I}$ is connected to at least two  checks in $\cal{E}$ and at most one checks in $\cal{O}$ and (b) No two checks of $\cal{O}$ are connected to a variable node outside $\cal{I}$.
\end{te}
\begin{proof}
See \cite{colwtthreepaper}.
\end{proof}
If the variable nodes corresponding to a trapping set are in error, then a decoder failure occurs. However, not all variable nodes corresponding to  trapping set need to be in error for a decoder failure to occur.

\begin{defn}\cite{chilappagarione} The minimal number of variable  nodes that have to be initially in error for the decoder to end up in the trapping set $\cal{T}$ will be referred to as {\it critical number} $m$ for that trapping set.\end{defn}
\begin{defn} \cite{colwtthreepaper} A set of variable nodes which if in error lead to a decoding failure is known as a \textit{failure set}.\end{defn}
\textit{Remarks}
\begin{enumerate}
\item To ``end up'' in a trapping set $\cal{T}$ means that, after a possible finite  number of iterations, the decoder will be in error, on at least one variable node from $\cal{T},$ at every iteration \cite{rich}. 
\item The notion of a failure set is more fundamental than a trapping set. However, from the definition, we cannot derive necessary and sufficient conditions for a set of variable nodes to form a failure set.
\item A trapping set is a failure set. Subsets of trapping sets can be failure sets. More specifically, for a trapping set of size $V$, there exists at least one subset of size equal to the critical number which is a failure set.
\item The critical number of a trapping set is not fixed. It depends on the outside connections of checks in $\cal{E}$. However, the maximum value of critical number of a $(V,C)$ trapping set is $V$.
\end{enumerate}

\section{Necessary and Sufficient Conditions to Correct Three Errors} \label{section3}
In this section, we establish the necessary and sufficient conditions for a column-weight-three code to correct three errors. We first illustrate three trapping sets and show that the critical number of these trapping sets is three thereby providing necessary condition to correct three errors. We then prove that avoiding structures isomorphic to these trapping sets in the Tanner graph is sufficient to guarantee correction of three errors.

\begin{figure*}[htb]
\centering
\subfigure[] % caption for subfigure a
{
    \label{sixcycle}

\includegraphics[width=0.18\textwidth]{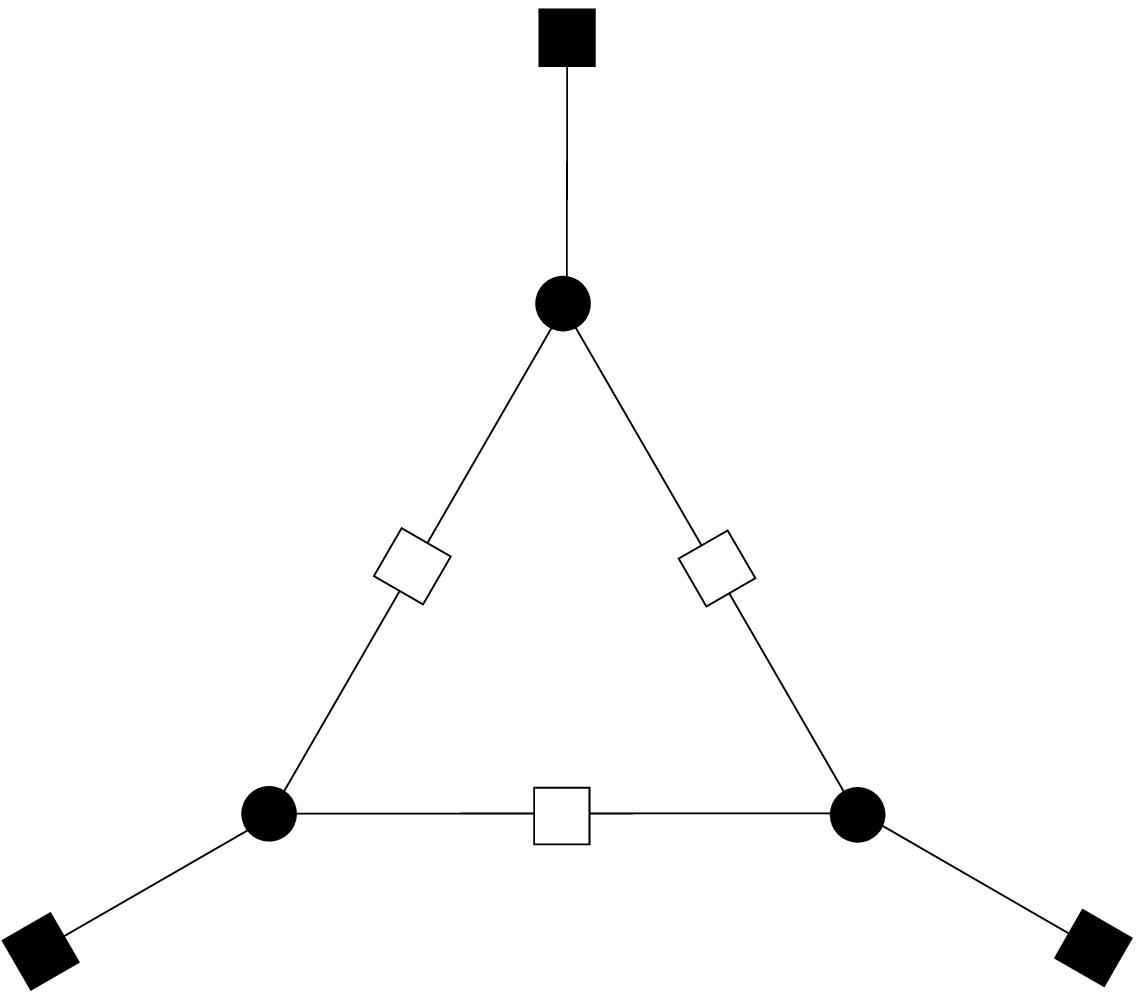}
}
\hspace{0.1\textwidth}
\subfigure[] % caption for subfigure a
{
    \label{53trappingset}

\includegraphics[width=0.3\textwidth]{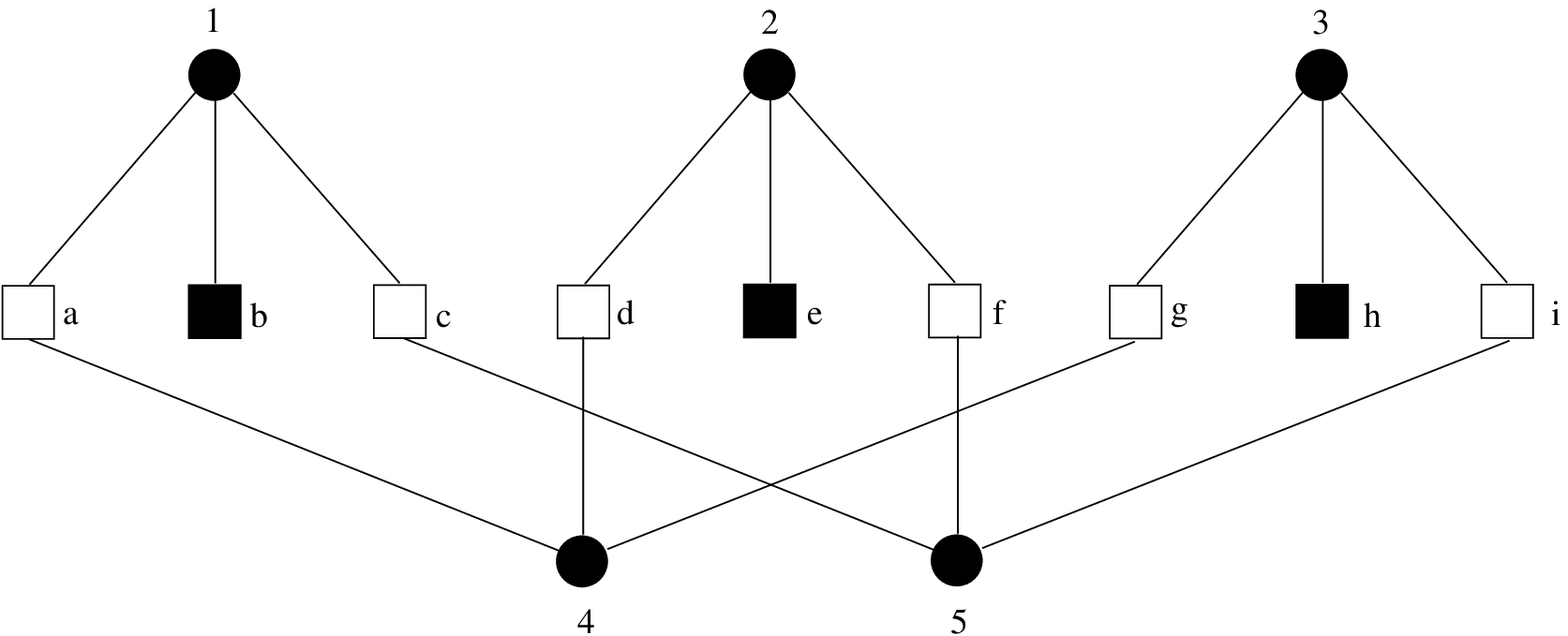}
}
\hspace{0.1\textwidth}
\subfigure[]  % caption for subfigure b
{
    \label{weight8codeword}

\includegraphics[width=0.25\textwidth]{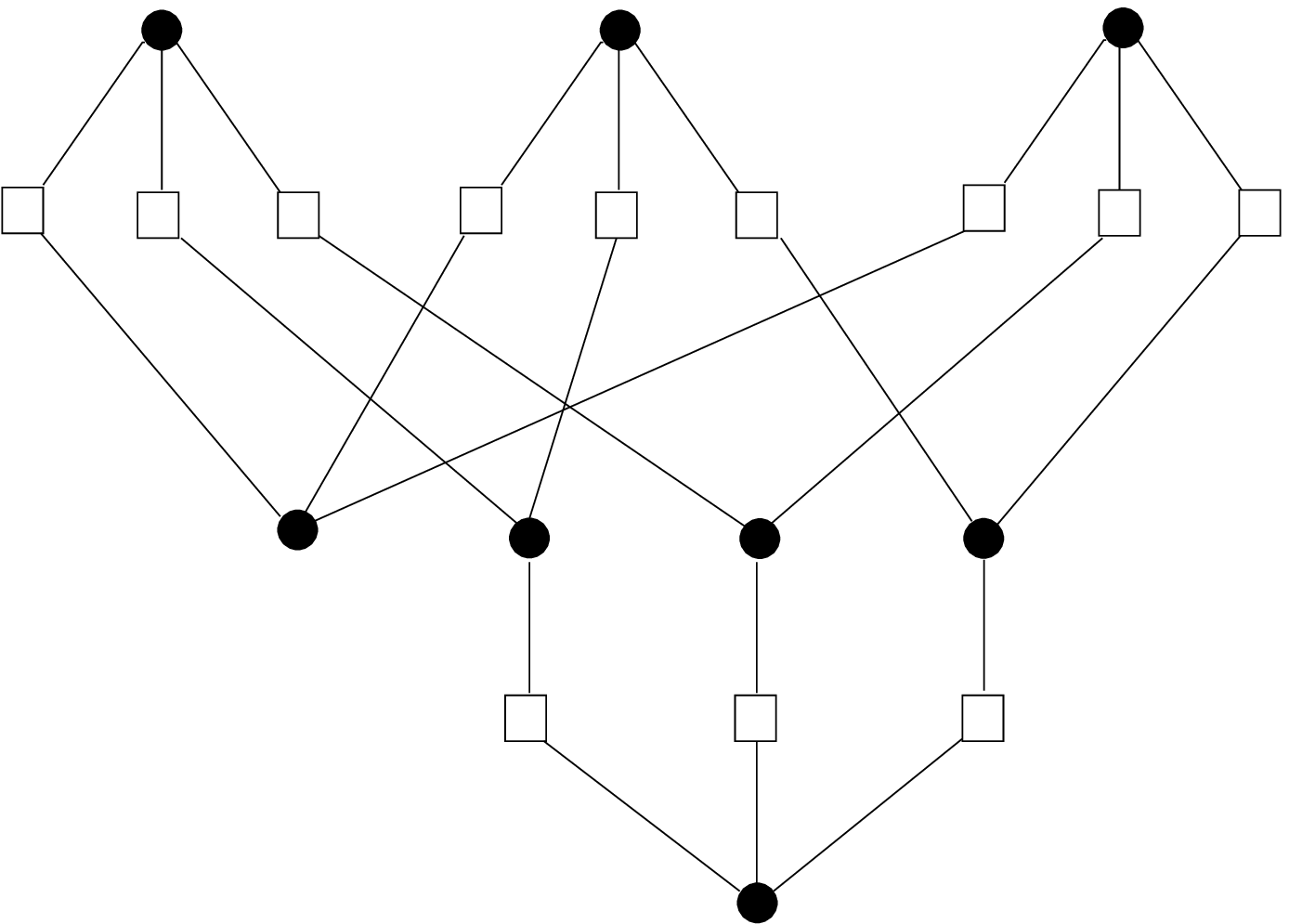}
}

\caption{Examples of trapping sets with critical number three \subref{sixcycle} a $(3,3)$ trapping set \subref{53trappingset} a $(5,3)$ trapping set and \subref{weight8codeword} an $(8,0)$ trapping set} \label{trappingsets}
\end{figure*}

Fig. \ref{trappingsets} shows three  subgraphs induced by different number of variable nodes. Let us assume that in all these induced graphs, no two odd degree checks are connected to a variable node outside the graph. By the conditions of Theorem \ref{thm1}, all these induced subgraphs are trapping sets. Fig. \ref{sixcycle} is a $(3,3)$ trapping set, Fig. \ref{53trappingset} is a $(5,3)$ trapping set and Fig. \ref{weight8codeword} is a $(8,0)$ trapping set. Note that a $(3,3)$ is isomorphic to a six cycle. and the $(8,0)$ trapping set is a codeword of weight eight.

\begin{figure*}[htb]
\centering
\subfigure[] % caption for subfigure a
{
    \label{config1}
\includegraphics[width=0.19\textwidth]{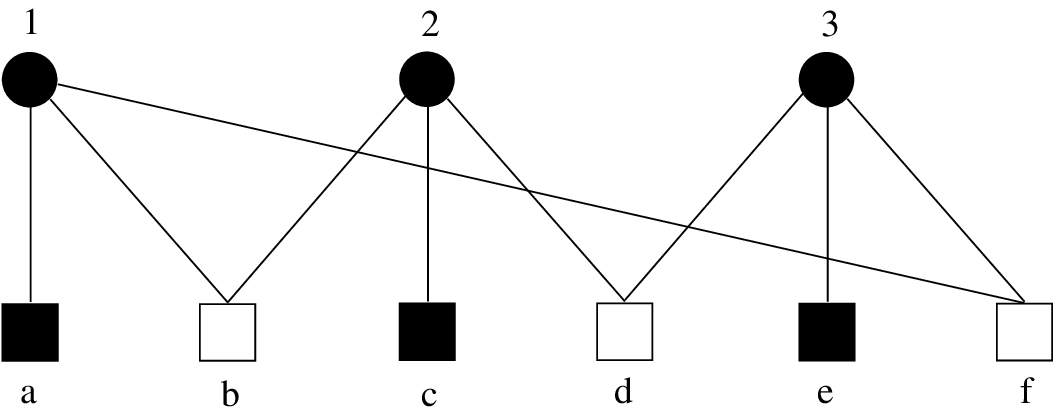}

}
\hspace{0.05\textwidth}
\subfigure[] % caption for subfigure a
{
    \label{config2}
\includegraphics[width=0.23\textwidth]{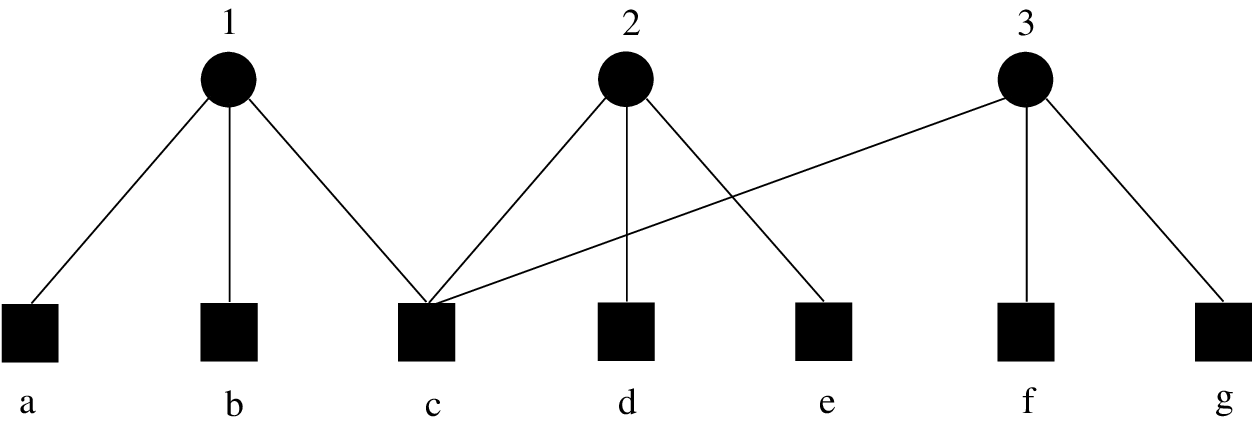}

}
\hspace{0.05\textwidth}
\subfigure[]  % caption for subfigure b
{
    \label{config3}

\includegraphics[width=0.23\textwidth]{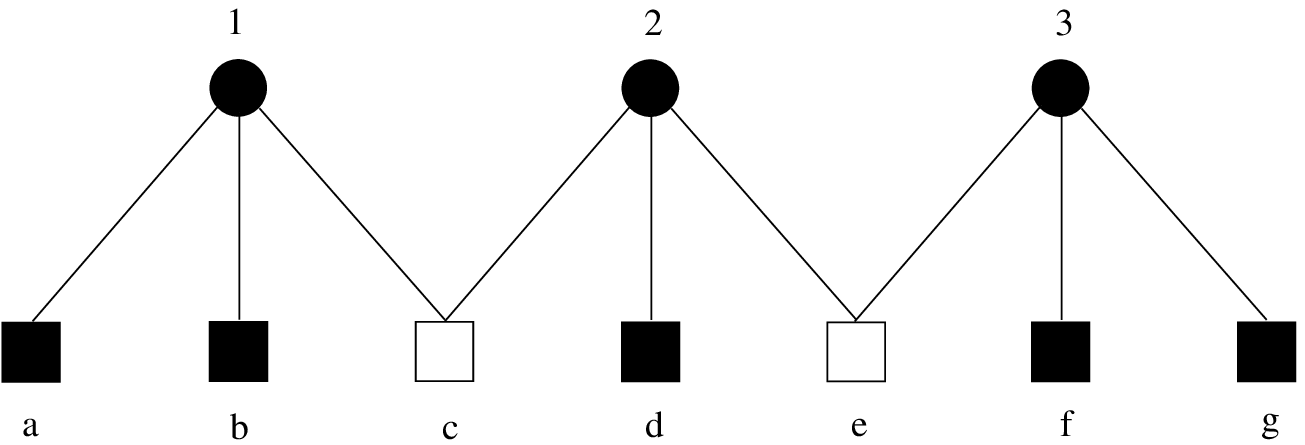}
}
\hspace{0.05\textwidth}
\subfigure[]  % caption for subfigure b
{
    \label{config4}

\includegraphics[width=0.25\textwidth]{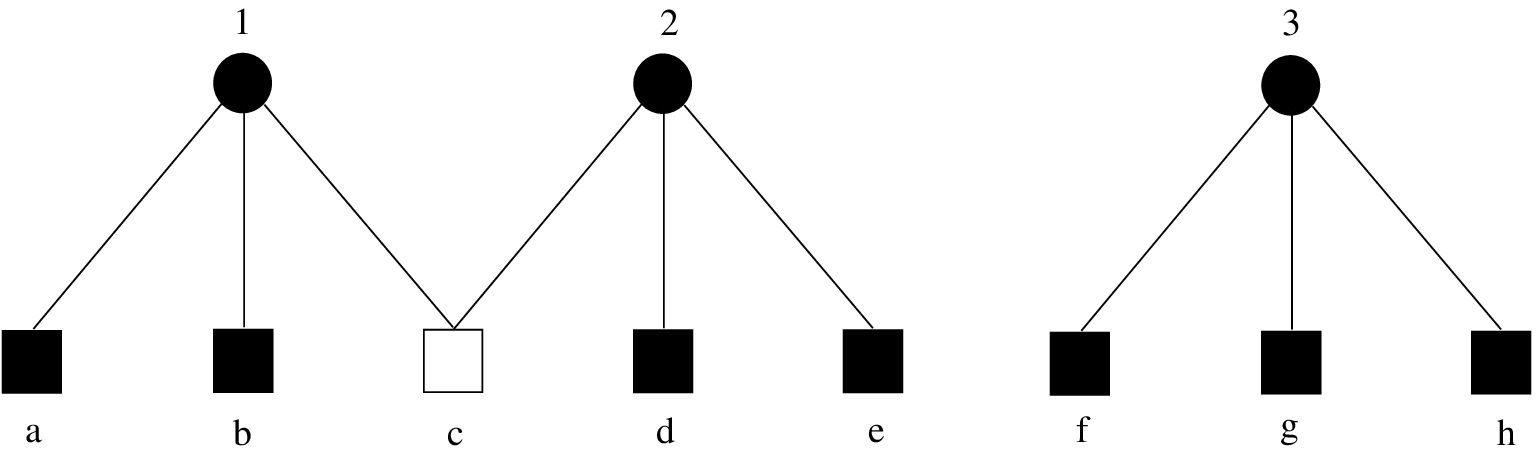}
}
\hspace{0.05\textwidth}
\subfigure[]  % caption for subfigure b
{
    \label{config5}
\includegraphics[width=0.28\textwidth]{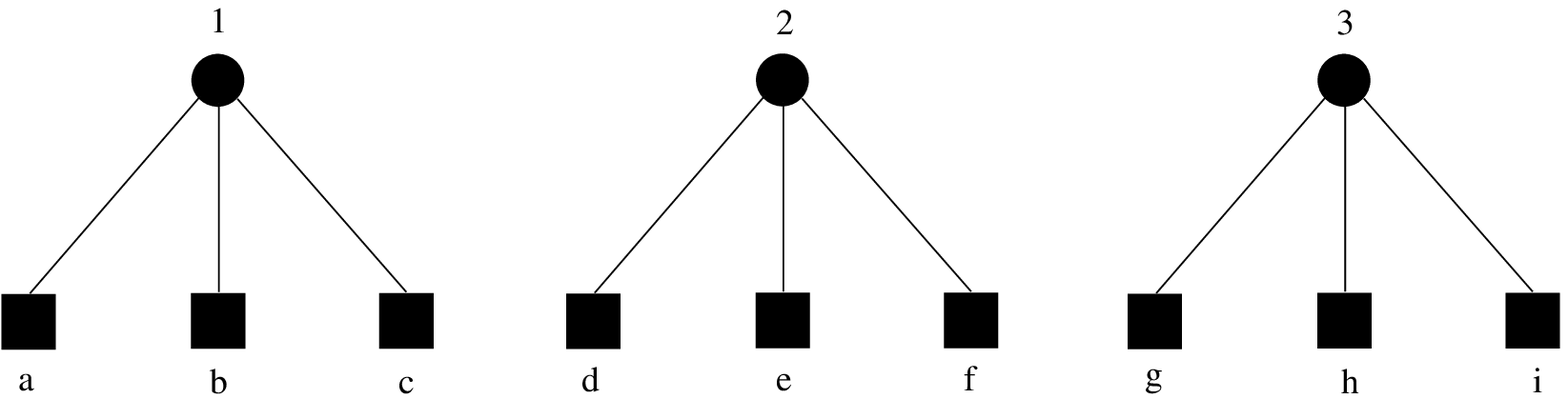}

}
 \caption{All the possible subgraphs that can be induced by three variable nodes in a column-weight-three code}\label{errorconfigs}
\end{figure*}

\begin{lema}
The critical number for $(3,3)$ trapping set is three. There exist $(5,3)$ and $(8,0)$ trapping sets with critical number three.
\end{lema}
\begin{proof}
For the $(3,3)$ trapping set, the result follows from definition. We omit the proof for $(5,3)$ and $(8,0)$ trapping sets due to space considerations. Detailed proofs can be found in the longer version of the paper \cite{jrnlpaperthreeerrors}.
\end{proof}
\begin{te}
To correct three errors in a column-weight-three LDPC code  by Gallager A algorithm, it is necessary to avoid $(3,3)$ trapping sets and $(5,3)$ and $(8,0)$ trapping sets with critical number three in its Tanner graph.
\end{te}
\begin{proof}
Follows from the above discussion.
\end{proof}
We now state and prove the main theorem.
\begin{te}
If the Tanner graph of a column-weight-three LDPC codes has girth eight and no set of variable nodes induces a subgraph isomorphic to $(5,3)$ trapping set or a subgraph isomorphic to $(8,0)$ trapping sets, then any three errors can be corrected using Gallager A algorithm.
\end{te}
\textit{Sketch of proof:} In a column-weight-three code three variable nodes can induce only one of the five subgraphs given in Fig. \ref{errorconfigs} and the proof proceeds by examining these subgraphs one at a time. The complete proof involves many arguments and here we just illustrate the methodology of the proof by considering two possible subgraphs. The proof for the remaining subgraphs appears in the longer version of the paper \cite{jrnlpaperthreeerrors}.
 
\textbf{Subgraph 1:} Since the girth of the code is eight, it has no six cycles and hence the configuration in Fig. \ref{config1} is not possible. 

\textbf{Subgraph 5:} The three variable nodes in error induce a subgraph as shown in Fig. \ref{config5}. In first half of first iteration  $1,2$ and $3$ send incorrect messages. In the second half of first iteration,  $a,b,c,d,e,f,g,h$ and $i$ send incorrect messages to neighboring variables except to $1,2$ and $3$. If there is no variable node which receives three incorrect messages, a valid codeword is reached after first iteration. On the contrary, assume there exists a variable node, say $4$, which receives three incorrect messages (w.l.o.g. we can assume that $4$ is connected to $a,d$ and $g$). Also, there cannot be two such variable nodes as that would introduce a six cycle or a graph isomorphic to $(5,3)$ trapping set. Also, there can be at most three variable nodes which receive two incorrect messages, say, $5,6$ and $7$. Let the other checks connected to these variables be $j,k$ and $l$ respectively. In the first half of second iteration, $1,2$ and $3$ send all correct messages, $4$ sends all incorrect messages, $5,6,7$ send incorrect messages to $j,k$ and $l$ respectively. In second half of second iteration, $a,d,g$ send incorrect messages to their neighbors except to $4$. $j,k$ and $l$ send incorrect messages to neighboring variables except to $5,6$ and $7$. There cannot be a variable node which is connected to one check from $\{j,k,l\}$ and to one check from $\{a,d,g\}$. Also, there cannot be a variable node which is connected to all the three checks $j,k$ and $l$ as this would introduce a graph isomorphic to $(8,0)$ trapping set. However, there can be at most two variable nodes which receive two incorrect messages from the checks $j,k$ and $l$, say $8$ and $9$. Let the other checks connected to $8$ and $9$ be $m$ and $p$. At the end of second iteration, $1,2$ and $3$ receive one incorrect message, $8$ and $9$ receive two incorrect messages. In the first half of third iteration, $1,2$ and $3$ send two incorrect messages each, $8$ and $9$ send one incorrect message each. In the second half of third iteration, $b,c,e,f,h$ and $i$ send incorrect messages to their neighbors except to $1,2$ and $3$. $m$ and $p$ send incorrect messages to their neighbors except to $8$ and $9$. It can be shown that there cannot exist a variable node which receives three incorrect messages. At the end of third iteration, $1,2$ and $3$ receive all correct messages and no variable node receives all incorrect messages. So, if a decision is made, a valid codeword is reached and decoder is successful.

\textit{Remark:} It is worth noting that the complete proof is more involved than the proofs which use expansion arguments. However, the result is also more precise and holds for codes of small lengths.  
\section{Numerical Results}\label{section4}
In this section, we describe a technique to construct codes which can correct three errors. Codes capable of correcting a fixed number of errors show superior performance on the BSC at low values of probability of transition $\alpha$. This is because the slope of the FER curve is related to the minimum critical number \cite{isitpaper}. A code which can correct $i$ errors has minimum critical number $i+1$ and the slope of FER curve is $i+1$. We restate the arguments from \cite{isitpaper} to make this connection clear. 

Let $\alpha$ be the transition probability of BSC and $c_k$ be number of configurations of received bits for which $k$ channel  errors lead to codeword (frame) error. The frame error rate (FER) is given by:

$$FER(\alpha)=\sum_{k=i}^n c_k\alpha^k(1-\alpha)^{(n-k)}$$  where $i$ is the minimal number of channel errors that can lead to a decoding error (size of instantons) and $n$ is length of the code.

On a semilog scale the FER is given by the expression

\begin{eqnarray}
\log(FER(\alpha))=\log \big(\sum_{k=i}^n c_k\alpha^k(1-\alpha)^{n-k}\big)  \nonumber \\
=\log(c_i)+i\log(\alpha)+\log((1-\alpha)^{n-i}) \nonumber \\
+\log\left(1+\frac{c_{i+1}}{c_i}\alpha(1-\alpha)^{-1}+\ldots+\frac{c_{n}}{c_i}\alpha^{n-i}(1-\alpha)^{-i}\right) \nonumber
\label{other}
\end{eqnarray}

In the limit $\alpha \rightarrow 0$ we note that
$$ \lim_{\alpha \rightarrow 0}
\Big[\log((1-\alpha)^{n-i})\Big]=0 $$
and
 $$ \lim_{\alpha
\rightarrow 0} \Big[\log \Big(1+
\frac{c_{i+1}}{c_i}\alpha(1-\alpha)^{-1}
\ldots+\frac{c_{n}}{c_i}\alpha^{n-i}(1-\alpha)^{i-n}\Big)
\Big]\!\!=0
$$
So, the behavior of the FER curve for small $\alpha$ is dominated by
$$\log(FER(\alpha)) \approx \log(c_i)+i\log(\alpha)$$

The $\log(FER)$ vs $\log(\alpha)$ graph is close to a straight line with slope equal to $i,$ the minimal critical number. If two codes $C_1$ and $C_2$ have minimum critical numbers $i_1$ and $i_2,$ such that $i_1 > i_2,$ then the code $C_2$ will perform better than $C_1$ for small enough $\alpha,$ independent of the number of trapping sets.

From the discussion in Section \ref{section3} and Section \ref{section4}, it is clear that for a code to have a FER curve with slope at least $4$, the corresponding Tanner graph should not contain the trapping sets shown in  Fig. \ref{trappingsets} as subgraphs. We now describe a method to construct such codes. The method can be seen as a  modification of the PEG construction technique used by Hu \textit{et al.} \cite{peg}. The algorithm is as follows:

\begin{algorithm}
 \KwData{The set of $n$ variable nodes ($V$) and $m$ check nodes ($C$). The column weight of the code ($\gamma$)}
 \KwResult{ Code with column weight $\gamma$}
 \For{$j = 1$ to $n$}{
 		\For{$k = 1$ to $\gamma$}{
 				\eIf{$k = 1$}{
 						Connect the $k^{th}$ edge of variable node $j$ to the check node with the smallest positive degree.
 				}
 				{
 						Expand the tree rooted at node $j$ to a depth of $6$.\\
 						Assimilate all check nodes which do not appear in the tree into $C_{j,\overline{T}}$, the set of candidates for connecting variable node 								$j$ to.\\
 						\While{$k^{th}$ edge is not found}{
 								Find the check node $c_i$ in $C_{j,\overline{T}}$ with the lowest degree. If connecting $c_i$ to variable node $j$ does not create a 										$(5, 3)$ trapping set, set this as the $k^{th}$ edge. If it does, remove $c_i$ from $C_{j,\overline{T}}$.
 						}
 				}
 		}
 }
\end{algorithm}

Note that checking for a graph isomorphic to $(8,0)$ trapping set at every step of code construction is computationally complex. Since, the PEG construction empirically gives good codes, it is unlikely that it introduces a weight-eight codeword. However, once the graph is grown fully, it can be checked for the presence of weight-eight codewords and these can be removed by swapping few edges. 

\begin{figure}

\includegraphics[width=3in]{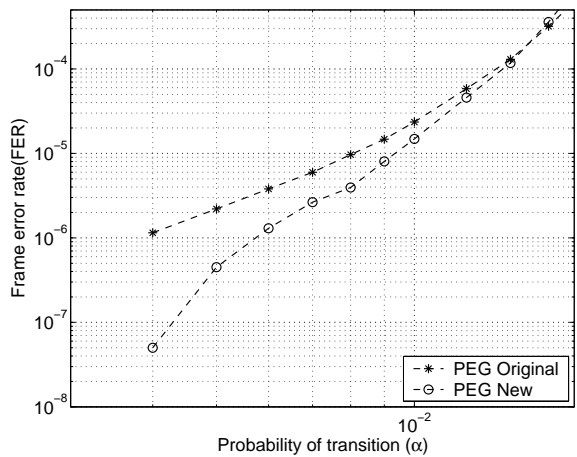}
\caption{Performance comparison of original PEG and the new PEG code}
\label{pegnewvsold}
\end{figure}

Using the above algorithm, a column-weight-three code with $504$ variable nodes and $252$ check nodes was constructed. The code has slight irregularity in check degree. There is one check node degree five and one check node with degree seven, but the majority of them have degree six. The code has rate 0.5. In the algorithm, we restrict maximum check degree to seven. The performance of the code on BSC is compared with the PEG code of same length. The PEG code is empirically the best known code at that length on AWGN channel \cite{mackayswebsite}. However, it has fourteen $(5,3)$ trapping sets. Fig. \ref{pegnewvsold} shows the performance comparison of the two codes. As can be seen, the new code performs better than the original PEG code at small values of $\alpha$.
\section{Conclusion}\label{section5}
In this paper, we have given conditions for a column-weight-three code to correct three errors. Since, the check degree does not play any part in the proof, it follows that the result is independent of code rate. A direction for future work is extending the analysis to more number of errors and higher column weight codes. Preliminary investigation shows a lot of promise. The complexity of the proof, even in the case of three errors, suggests that solving the problem for an arbitrary number of errors will be a challenge. On the code construction front, we have shown that avoiding trapping sets with minimum critical number is the criterion to suppress error floor. However, the conditions for correcting more errors could be more complicated thereby increasing the complexity of code construction. Deriving bounds on lengths and minimum distance of codes which avoid certain structures  also need to be investigated.

\section*{Acknowledgment}
This work is funded by NSF under Grant CCF-0634969 and INSIC-EHDR program.

\end{document}